\documentclass[prl,twocolumn,showpacs,amsmath,amssymb,epsfig]{revtex4-1} 
 
\usepackage{graphicx}
\usepackage{dcolumn}
\usepackage{bm}
\usepackage{mathrsfs}

\usepackage{natbib}
\usepackage[text={7in,9.5in},centering]{geometry}

\usepackage{hyperref}
\usepackage{url}

\usepackage{epsfig}
\usepackage{amsmath}
\usepackage{amssymb}
\usepackage{textcomp}

\usepackage{color}
\usepackage{float}

\usepackage{stackengine}
\usepackage{verbatim}
\begin{document}

\title{Observation of the X17 anomaly in the decay of the Giant Dipole Resonance of $^8$Be}
\author{A.J. Krasznahorkay}
\email{kraszna@atomki.hu}
\author{A.~Krasznahorkay}
\altaffiliation{Currently working at CERN, Geneva, Switzerland}

\author{M. Csatl\'os}
\author{L. Csige}
\author{J. Tim\'ar}
\author{M. Begala}
\author{A. Krak\'o}
\author{I. Rajta}
\author{I. Vajda}
\affiliation{Institute for Nuclear Research  (ATOMKI),
  P.O. Box 51, H-4001 Debrecen, Hungary}
\author{N.J. Sas}
\affiliation{University of Debrecen, 4010 Debrecen, PO Box 105, Hungary}

\begin{abstract}
Angular correlation spectra of $e^+e^-$ pairs produced in the
$^{7}$Li($p$,$\gamma$)$^{8}$Be nuclear reaction were studied at a
proton beam energy of $E_p$~=~4.0~MeV, which corresponds to the
excitation energy of the Giant Dipole Resonance (GDR) in $^8$Be. The
spectra measured show a peak like anomaly at 120$^\circ$ and a broader
anomaly also above 140$^\circ$.  Both anomalies could consistently be
described by assuming that the same hypothetical X17 particle was
created both in the ground-state transition and in the transition
going to the broad ($\Gamma$=1.5~MeV), first excited state in $^8$Be.
The invariant mass of the particle, which was derived to be $m_Xc^2 =
16.95 \pm 0.48$(stat.)~MeV, agrees well with our previously published
values.
\end{abstract}


\maketitle

\section{Introduction}

We published very challenging experimental results in 2016 \cite{kr16}
indicating the electron-positron ($e^+e^-$) decay of a hypothetical
new light particle.  The $e^+e^-$ angular correlations for the
17.6~MeV and 18.15~MeV transitions in $^8$Be were studied and an
anomalous angular correlation was observed for the 18.15~MeV
transition \cite{kr16}. This was interpreted as the creation and decay
of an intermediate bosonic particle with a mass of
$m_{X}c^2$=16.70$\pm$0.35(stat)$\pm$0.5(sys)~MeV, which is now called
X17.

Our data were first explained with a vector gauge boson, X17 by Feng
and co-workers \cite{fe16,fe17,fe20}, which would mediate a fifth
fundamental force with some coupling to standard model (SM) particles.
The possible relation of the X17 boson to the dark matter problem
triggered an enormous interest in the wider physics community
\cite{ins}. New results will hopefully be published soon on the X17
particle from a few different experiments \cite{al23}.

We also observed a similar anomaly in $^4$He \cite{kr21}. It could be
described by the creation and subsequent decay of a light particle
during the proton capture process on $^3$H to the ground state of
$^{4}$He. The derived mass of the particle ($m_{X}c^2 = 16.94 \pm
0.12$(stat.)$\pm 0.21$(syst.)~MeV) agreed well with that of the
proposed X17 particle.

Recently, we have studied the E1 ground state decay of the 17.2 MeV
J$^\pi$ = 1$^-$ resonance in $^{12}$C \cite{kr22}.  The angular
correlation of the $e^+e^-$ pairs produced in the
$^{11}$B(p,$\gamma$)$^{12}$C reaction were studied at five different
proton energies around the resonance. The gross features of the
angular correlations can be described well by the Internal Pair
Creation (IPC) process following the E1 decay of the $1^-$ resonance.
However, on top of the smooth, monotonic distribution, we observed
significant peak-like anomalous excess around 155-160$^\circ$ at four
different beam energies. The $e^+e^-$ excess can be well-described by
the creation and subsequent decay of the X17 particle.  The invariant
mass of the particle was derived to be ($m_{X}c^2 = 17.03 \pm
0.11$(stat.)$\pm 0.20$(syst.)~MeV), in good agreement with our
previously published values.

However, despite the consistency of our observations, more
experimental data are needed to understand the nature of this
anomaly. For this reason, many experiments all over the world are in
progress to look for such a particle in different channels. Many of
these experiments have already put constraints on the coupling of this
hypothetical particle to ordinary matter. Others are still in the
development phase, but hopefully they will soon contribute to a deeper
understanding of this phenomenon as concluded by the community report
of the Frascati conference \cite{al23}.

Very recently, Barducci and Toni published an updated view on the
ATOMKI nuclear anomalies \cite{ba23}. They have critically re-examined
the possible theoretical interpretation of the observed anomalies in
$^8$Be, $^4$He and $^{12}$C anomalies in terms of a BSM boson X with
mass $\approx$17 MeV. Their results identify an {\it axial vector
  state} as the most promising candidate to simultaneously explain all
three anomalous nuclear decays, while the other spin/parity
assignments seems to be disfavored for a combined explanation.

At the same time, the NA62 collaboration was searching for K$^+$
decays to the $\pi^+e^+e^-e^+e^-$ final state and excluded the QCD
axion as a possible explanation of the 17 MeV anomaly \cite{NA23}.
Hostelt and Pospelov reanalysed some old pion decay constraints
\cite{po23}, ruled out the vector-boson explanations and set limits on
axial-vector
ones.

The aim of this paper is to use a simpler geometry of the spectrometer
to avoid non-trivial possible artefacts, which may be connected to the
spectrometer itself \cite{al21}.

With such a new spectrometer, we studied the X17 creation and the
$e^+e^-$ pair emission from the decay of the Giant Dipole Resonance
(GDR) \cite{fi76,sn86,ha01} excitations of $^8$Be.

\section{Experimental methods}

The experiments were performed in Debrecen (Hungary) at the 2 MV
Tandetron accelerator of ATOMKI, with a proton beam energy of E$_p$=
4.0 MeV.

Owing to the rather large width of the GDR($\Gamma$~=~5.3~MeV
\cite{fi76}), a 1 mg/cm$^2$ thick $^{7}$Li$_2$O target was used in
order to maximize the yield of the $e^+e^-$ pairs. The target was
evaporated onto a 10 $\mu$m thick Ta foil. The average energy loss of
the protons in the target was $\approx$100~keV.

$\gamma$ radiations were detected by a 3''x3'' LaBr3 detector
monitoring also any potential target losses. The detector was placed
at a distance of 25 cm from the target at an angle of 90 degrees to
the beam direction.

\begin{figure}[htb]
  \begin{center}
    \includegraphics[scale=0.4]{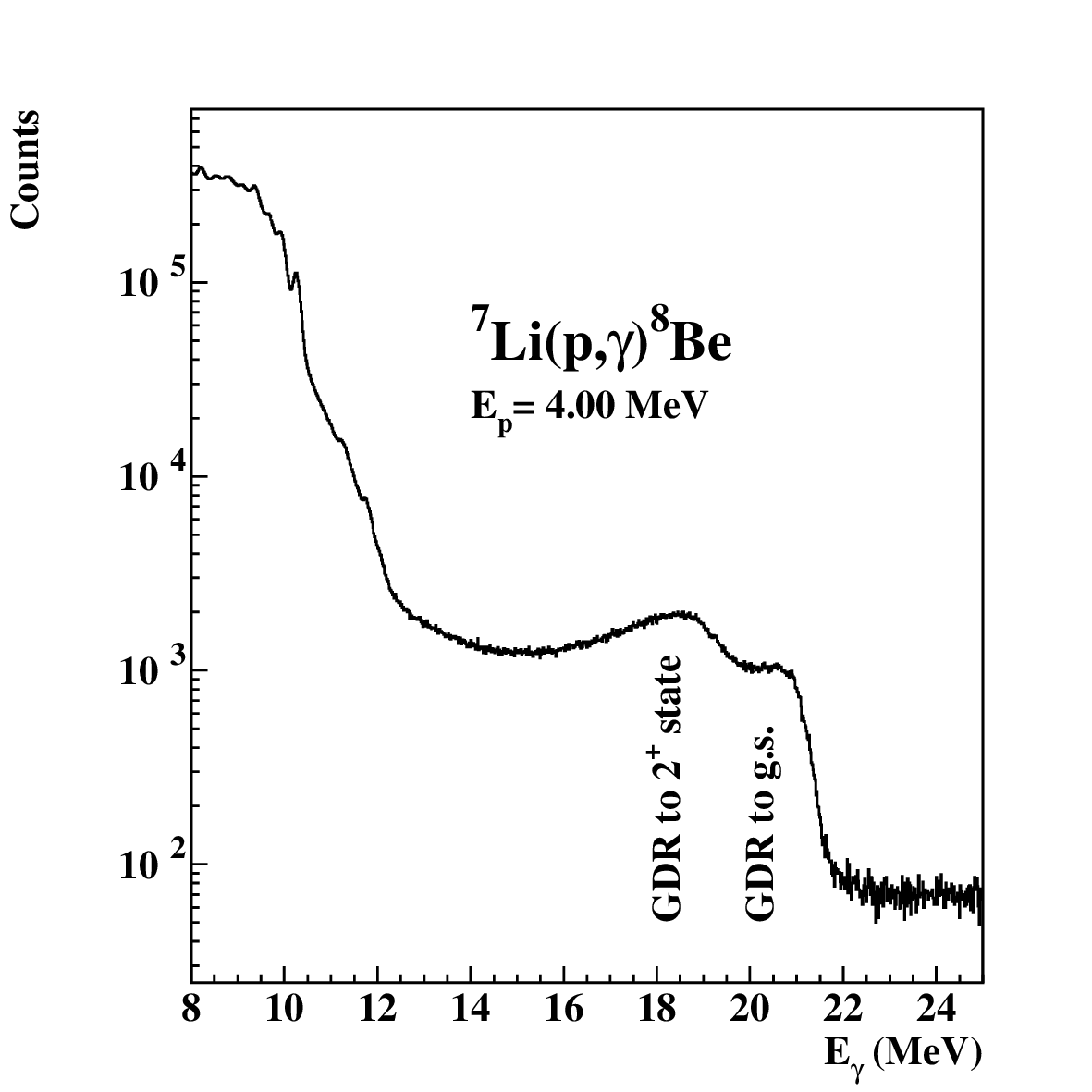}
  \end{center}
  \caption{Tipical $\gamma$-ray spectrum measured for the
    $^{7}$Li($p$,$\gamma$)$^{8}$Be nuclear reaction at $E_p$= 4.0 MeV}
  \label{fig:gamma}
\end{figure}

A typical $\gamma$ energy spectrum is shown in
Fig~\ref{fig:gamma}. The figure clearly shows the transitions from the
decay of GDR to the ground and first excited states in $^{8}$Be. The
cosmic ray background is also visible on the high energy side of the spectrum,
but it is reasonably low.

The intensity ratio of the peaks was found to be: I(GDR$\rightarrow $
g.s.)/I(GDR$\rightarrow 2_1^+$)=0.18$\pm$0.02 at E$_p$= 4.0 MeV
bombarding energy.

We used Double-sided Silicon Strip Detectors (DSSD) and plastic
scintillators as ``particle telescopes'' to determine the hit positions
and the energy of the electrons and positrons, respectively. In our
previous experiments, the spectrometers were built up of five and six
particle telescopes, both having different acceptances as a function
of the $e^+e^-$ correlation angle. A detailed description of the
spectrometers can be found in Ref.~\cite{kr21}. However, in the
present experiment, only two telescopes were used, placed at an angle
of 110$^\circ$ with respect to each other. The diameter of the carbon
fiber tube of the target chamber has been reduced from 70 mm to 48 mm
to allow a closer placement of the telescopes to the target.  Thus, we
could cover a solid angle around 110$^\circ$ with the two telescopes
much larger than with the previous setups. Also, in this setup the
efficiency function has only one maximum as a function of the $e^+e^-$
opening angle. This angular dependence can be simulated and calibrated
more reliably. Another advantage of this setup is that its sensitivity
to the cosmic background is significantly less. Since the vertical
angles of telescopes were
-35$^\circ$ and -145$^\circ$, the cosmic rays coming mostly vertically,
have a very small chance of firing both telescopes at the same time.

\begin{figure}[htb]
  \begin{center}
    \includegraphics[scale=0.4]{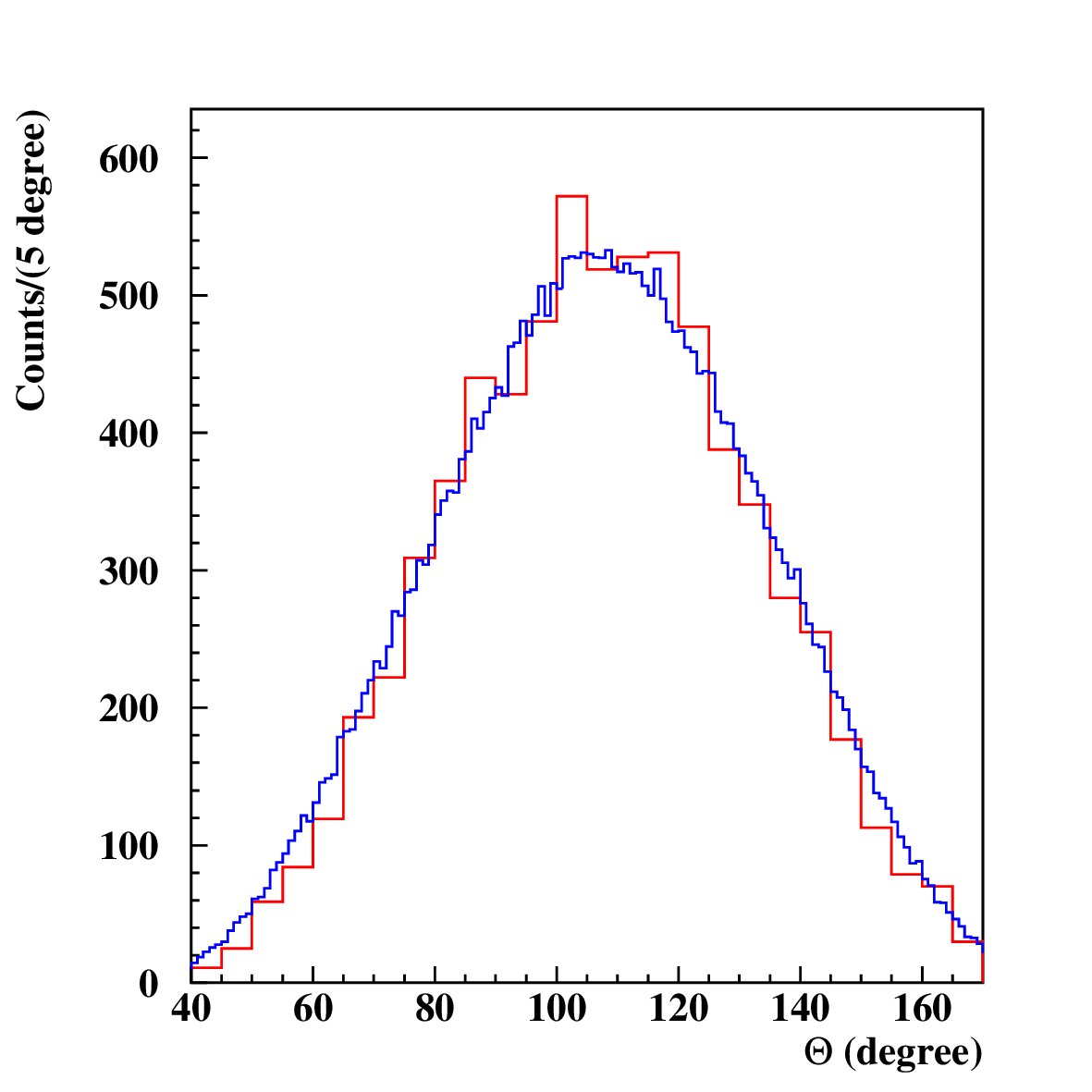}
  \end{center}
  \caption{Experimental acceptance of the spectrometer as a function of
    correlation angle ($\theta$) for consecutive, uncorrelated  e$^+$e$^-$
    pairs (red line histogram) compared with the results of the MC
    simulations (blue line histogram) as explained in the
    text.}
  \label{fig:acceptance}
\end{figure}

The energy calibration of the telescopes, the energy and position
calibrations of the DSSD detectors, the Monte Carlo (MC) simulations
as well as the acceptance calibration of the whole $e^+e^-$
coincidence pair spectrometer were explained in Ref. \cite{kr21}.
Good agreement was obtained between the experimental acceptance and
the results of the MC simulations, as presented in
Fig.~\ref{fig:acceptance}. Due to the very tight geometry, the DSSD
position data and therefore the $e^+e^-$ angular distribution
experiences an enhanced dependence on the beam spot size and
position. According to previous measurements and MC simulations of the
present setup we could take into account this effect properly.

At the proton energy of E$_p$= 4.0 MeV, the (p,n) reaction channel is
open (E$_{thr}$= 1.88 MeV) and generated neutrons and low-energy
$\gamma$ rays with a large cross section.  (Other reaction channels
are also open, but their cross sections are much smaller and their
influence on our experiment is much weaker.)  The maximum neutron
energy E$_n=1.6$~MeV, which induces only a 300 keV electron equivalent
signal in the plastic scintillator due to the quenching effect. Such a
small signal fell well below the CFD thresholds that we used.

The low-energy neutrons did not produce any measurable signal in the
DSSD detectors either since the maximum energy that can be transferred
in elastic scattering on Si atoms is only $\approx$50 keV, which is
below the detection threshold.

\begin{figure}[htb]
\begin{center}
\includegraphics[scale=0.4]{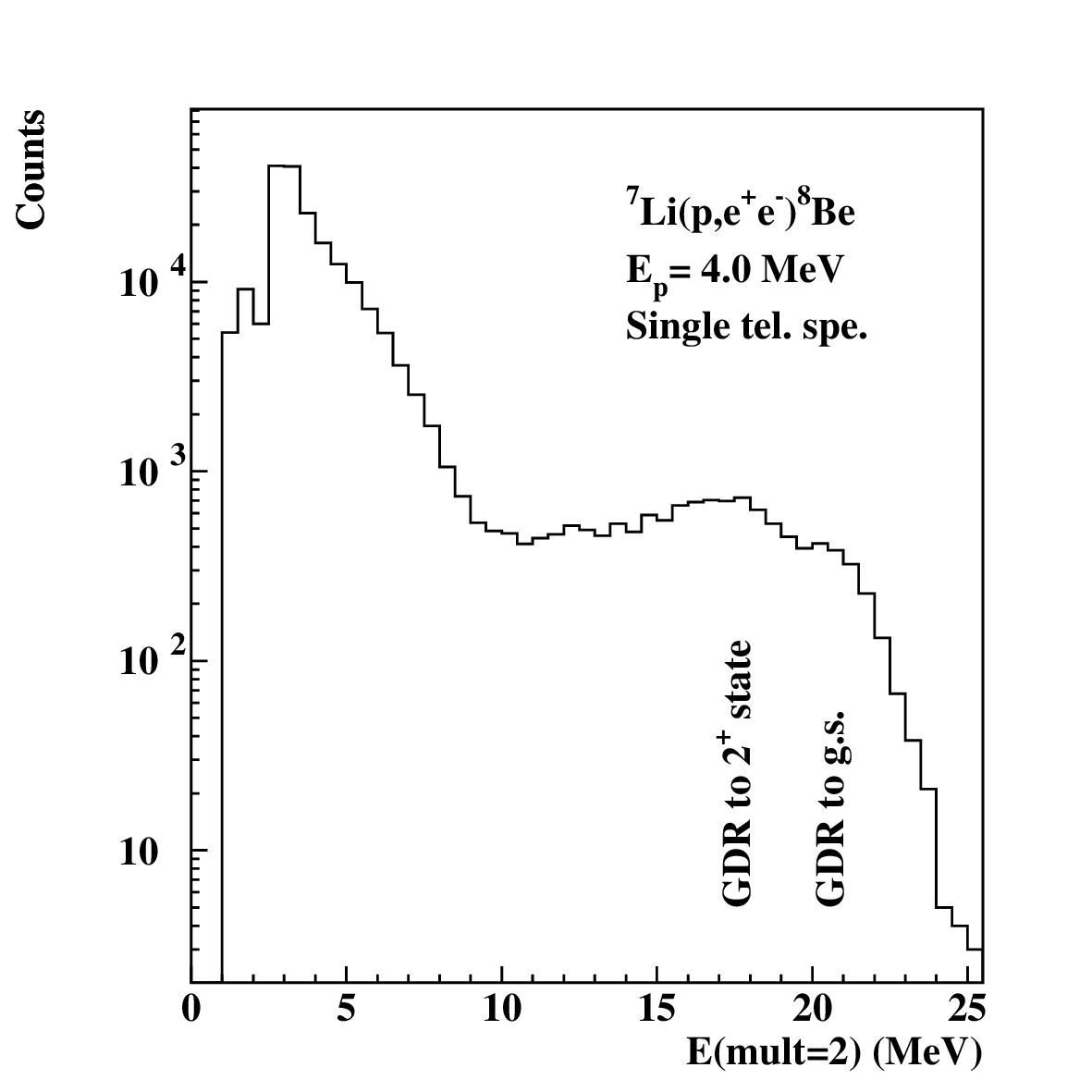}
\end{center}
\caption{Total energy spectrum of the $e^+e^-$-pairs from the
  $^{7}$Li(p,$e^+e^-$)$^{8}$Be nuclear reaction measured in telescope 1
  by requiring multiplicity=2 in their corresponding DSSD detector.}
\label{enesum}
\end{figure}

A single energy spectrum measured by the scintillators and gated by
``multiplicity=2'' events in the DSSD detector, which means that both
the electron and positron coming from the internal pair creation are
detected in the same telescope, is used for energy calibration. Such a
calibration spectrum is shown in Fig.~\ref{enesum} for telescope 1.

As shown, the energy resolution for the
ground-state transition is reasonably good ($\approx 14\%$).
The intensity ratio of the GDR to ground state and the GDR to the 2$_1^+$ state
is determined to be:
I(GDR$\rightarrow$g.s.)/I(GDR$\rightarrow$2$_1^+$)=0.25$\pm 0.03$.

\section{Experimental results}

\begin{figure}[htb]
\begin{center}
\includegraphics[scale=0.4]{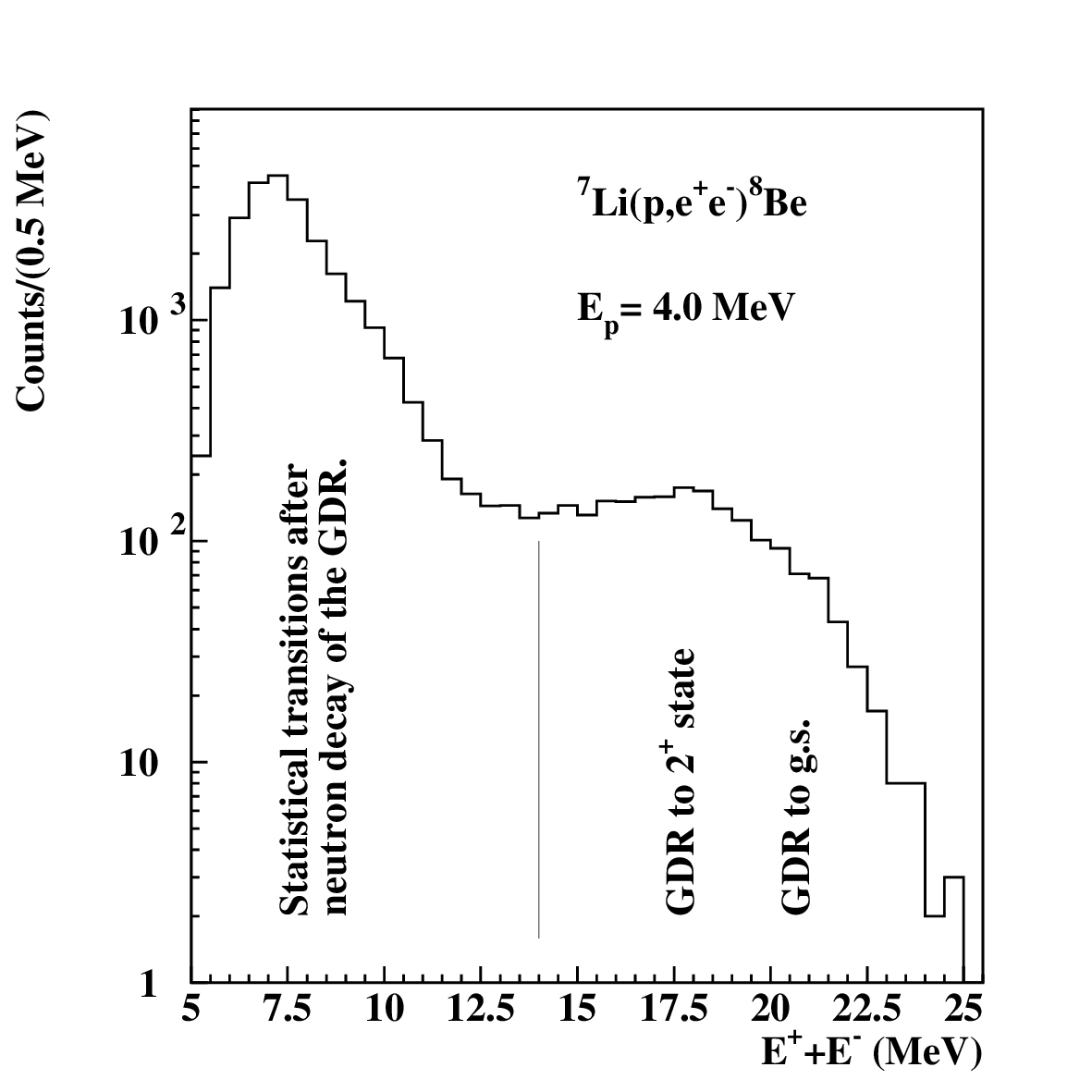}
\end{center}
\caption{Total energy spectrum of the $e^+e^-$-pairs from the
  $^{7}$Li(p,$e^+e^-$)$^{8}$Be nuclear reaction.}
\label{Fig:sume}
\end{figure}

Unfortunately, the gain of the PMT connected to the second plastic
scintillator was less stable than the first one and its energy
resolution was somewhat worse. This is represented by the worse
resolution of the energy sum spectrum of the two telescopes as shown
in Fig.~\ref{Fig:sume}.

The angular correlation spectra of the $e^+e^-$ pairs for the
different energy sum regions were then obtained for symmetric
$-0.5\leq\epsilon \leq 0.5$ pairs, where the energy asymmetry
parameter $\epsilon$ is defined as $\epsilon=(E_1-E_2)/(E_1 + E_2)$,
where $E_1$ and $E_2$ denote the kinetic energies of the leptons
measured in telescope 1 and telescope 2, respectively.

\begin{figure}[htb]
    \begin{center}
\includegraphics[scale=0.4]{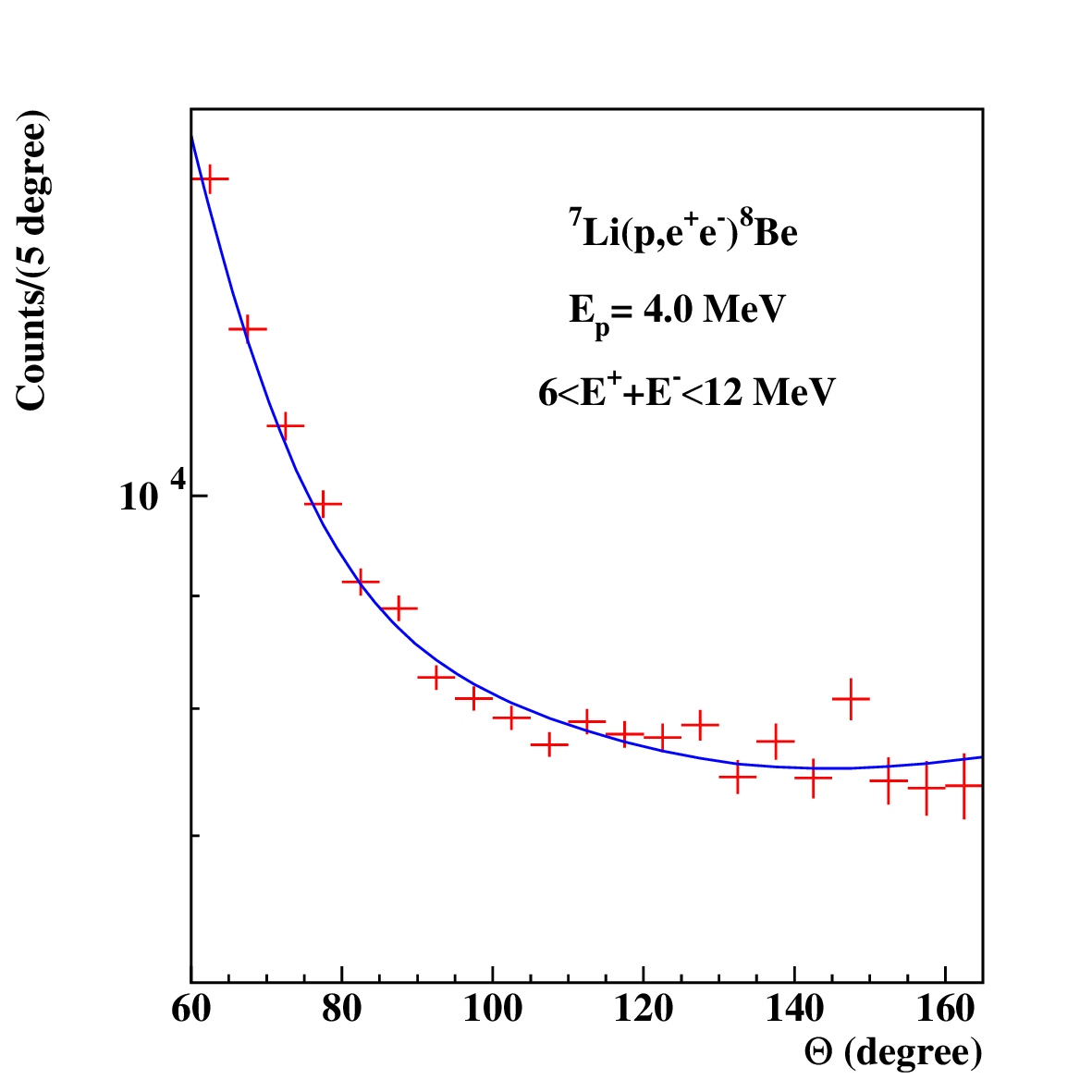}
    \end{center}
\caption{Experimental angular correlations of the $e^+e^-$ pairs measured in
  the $^{7}$B(p,$e^+e^-$)$^{8}$Be reaction at E$_p$=4.0 MeV for low-energy
  (E$^+$+E$^-$ $\leq$14 MeV)
  transitions.}
     \label{Fig:ang-low}
\end{figure}

The angular correlation gated by the low energy-sum region (below 14
MeV), as marked in Fig.~\ref{Fig:sume}, is shown in
Fig.~\ref{Fig:ang-low}. The measured counts were corrected for the
acceptance obtained from the raw data collected for the whole
experiment in the similar way as described previously \cite{kr21}.  It
is a smooth distribution without showing any anomalies. It could be
described by assuming E1 + M1 multipolarities for the IPC process and
a constant distribution, which may originate from cascade transitions
of the statistical $\gamma$ decay of the GDR appearing in real
coincidence. In such a case, the lepton pair may come from different
transitions, and thus their angles are uncorrelated.  This smooth
curve reassured us that we were able to accurately determine the
efficiency of the spectrometer.

\begin{figure}[htb]
    \begin{center}
\includegraphics[scale=0.4]{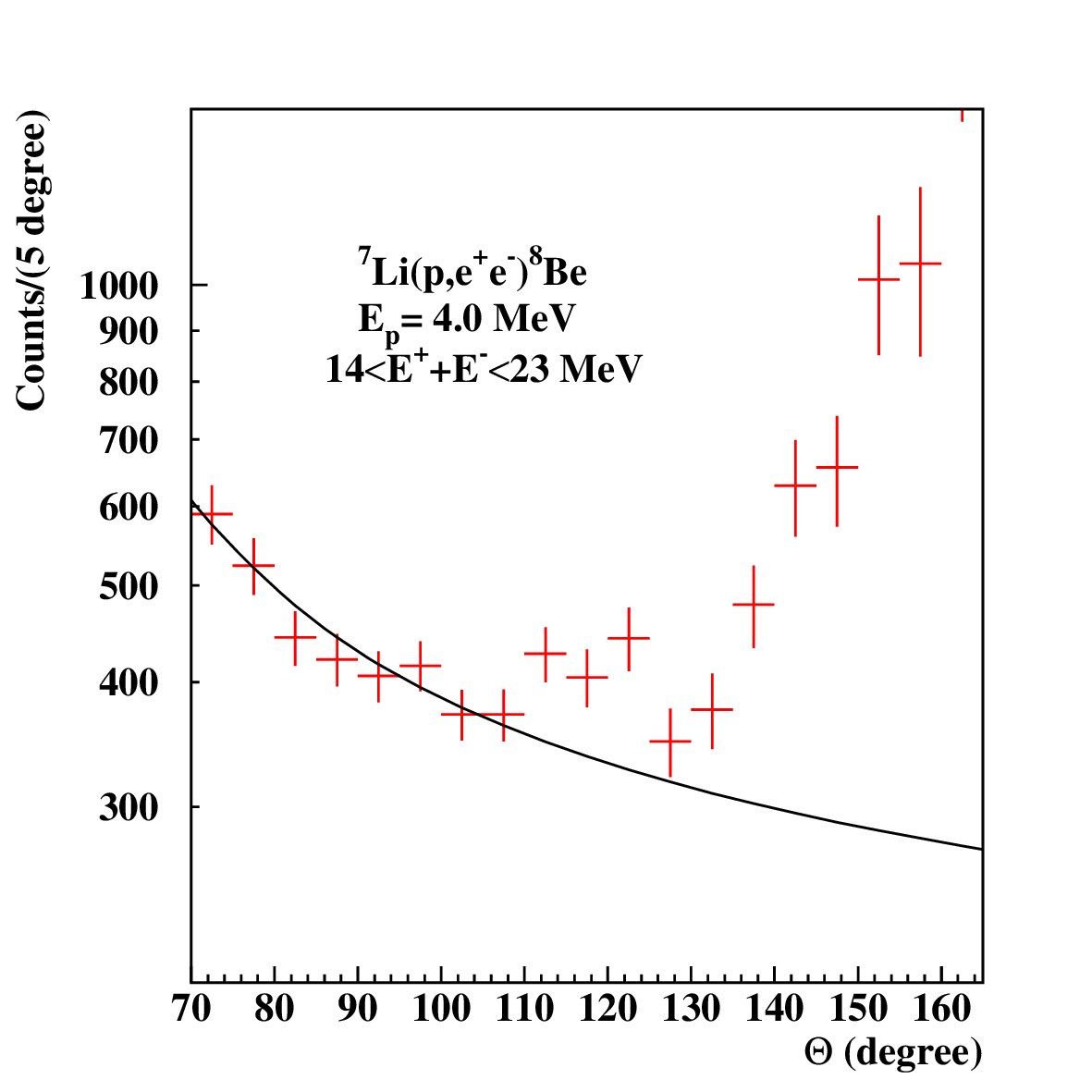}
    \end{center}
\caption{Experimental angular correlations of the $e^+e^-$ pairs measured in
  the $^{7}$Li(p,$e^+e^-$)$^{8}$Be reaction at E$_p$=4.0 MeV at the vicinity of
  the GDR. See explanation in the text.}
     \label{Fig:ang-gdr}
\end{figure}

The angular correlation of the $e^+e^-$ pairs gated by the GDR energy
region (above 14 MeV), as marked in Fig.~\ref{Fig:sume}, is shown
Fig.~\ref{Fig:ang-gdr}.

The experimental data corrected for the acceptance of the spectrometer
is shown as red dots with error bars. The simulated angular
correlation for the E1 internal pair creation is indicated as a black
curve. Significant deviations were observed. First of all, a peak-like
deviation at 120$^\circ$, but also an even stronger deviation at
larger angles.

The measured angular correlation was fitted from 70 degrees to 160
degrees with the sum of simulated E1, M1 and X17 contributions
calculated for both the GDR to ground state and for the GDR to 2$_1^+$
state transitions. The simulations concerning the decay of the X17
boson in the transition to the ground state of $^8$Be were carried out
in the same way as we did before \cite{kr16,kr21,kr22} and could
describe the anomaly appearing at around 120$^\circ$.

However, based on Figures \ref{enesum} and \ref{Fig:sume} and previous
measurements \cite{fi76}, the $\gamma$-decay of GDR to the first
excited state is much stronger than its decay to the ground state.
According to that, we assumed that the X17 particle was created also
in the decay of GDR to the ground state and to the first excited
state.  Based on the energy of that transition (17.5 MeV), we would
expect a peak around 150 degrees. However, the first excited state is
very broad ($\Gamma$=1.5 MeV), so the shape of the expected anomaly is
significantly distorted.  The simulations were then performed as a
function of the X17 mass from 10 MeV/c$^2$ to 18 MeV/c$^2$ for both
transitions.

To derive the invariant mass of the decaying particle, we carried out
a fitting procedure for both the mass value and the amplitude of the
observed peaks. The fit was performed with RooFit
\cite{Verkerke:2003ir} in a similar way as we described before
\cite{kr21,kr22}.

The experimental $e^+e^-$ angular correlation was fitted  with the following
intensity function
(INT):

\begin{equation}
\label{eq:pdf}
\begin{split}
INT&(e^+e^-) = \\
 &N_{E1} * PDF(E1) + N_{M1} * PDF(M1) + \\
 &N_{Sig} * \alpha_{ground}* PDF(sigground) + \\
 &N_{Sig} * (1 - \alpha_{ground}) * PDF(sig2plus)\ ,
\end{split}
\end{equation}

\noindent
where where PDF(X) represents the MC-simulated probability density
functions. $PDF(E1), PDF(M1)$ were simulated for Internal Pair
Creation having electromagnetic transitions with E1 and M1
multipolarity. $PDF(sigground), PDF(sig2plus)$ were simulated for the
two-body decay of an X17 particle as a function of its mass created in
the GDR to the ground state and GDR to $2_1^+$ transitions,
respectively.  $N_{E1}$, $N_{M1}$, and $N_{Sig}$ are the fitted
numbers of background and signal events, respectively.
$\alpha_{ground}$ is the fraction of X17 decays detected in the GDR to
ground state transition, with respect to the total number of detected
X17 decays. We assumed the same mass for the X17 particle created in
the two transitions.  The result of the fit is shown in
Fig.~\ref{Fig:ang-gdr} together with the experimental data.

\begin{figure}[htb]
    \begin{center}
\includegraphics[scale=0.4]{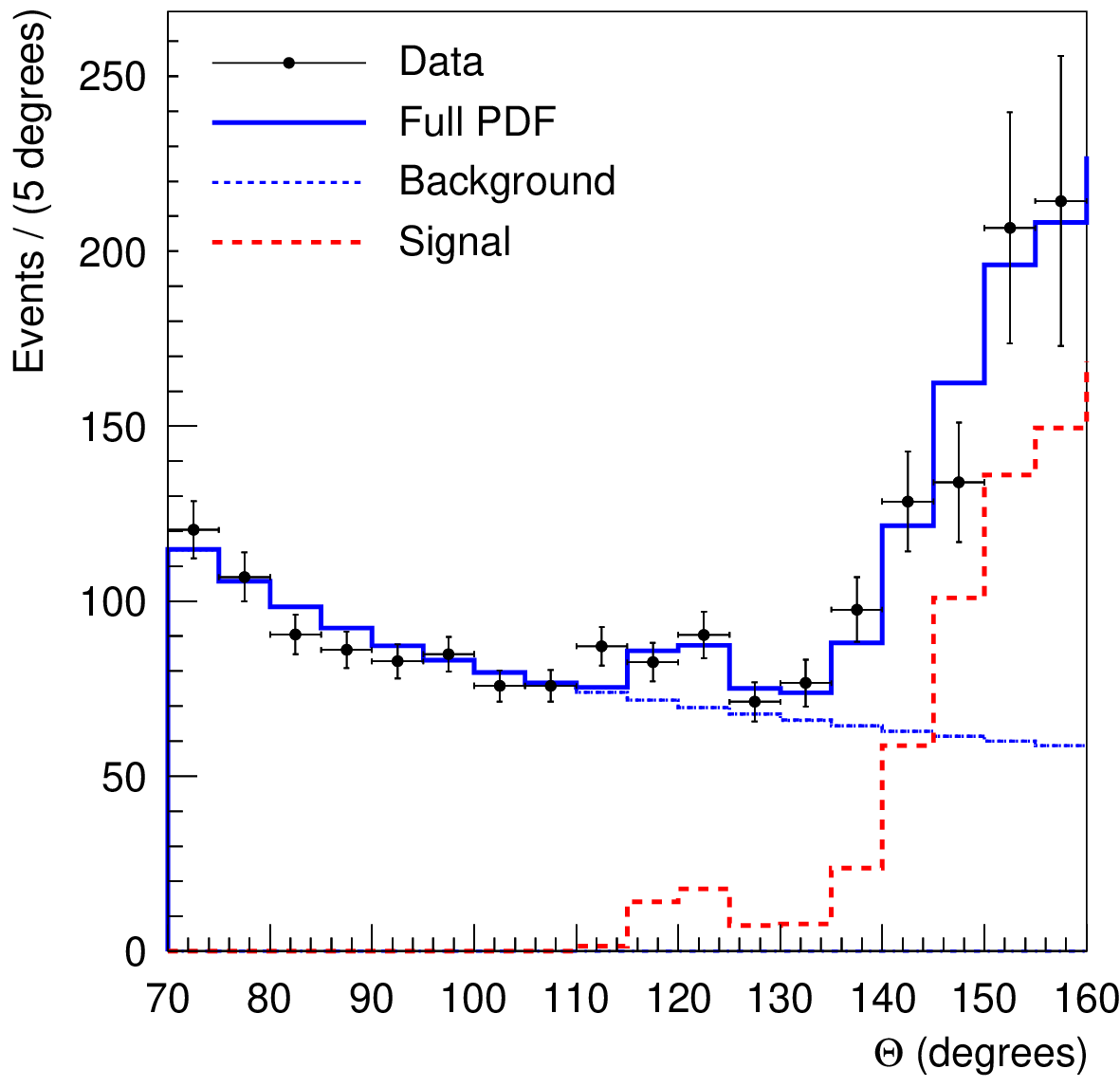}
    \end{center}
\caption{Experimental angular correlations of the $e^+e^-$ pairs
  fitted by the contributions from the E1 IPC and from the
  contributions coming from the $e^+e^-$ decay of the X17 particle.}
     \label{Fig:ang-fit}
\end{figure}

As shown in Fig.~\ref{Fig:ang-fit}, the simulation can describe the
experimental distributions from $\Theta = 70^\circ$ to 160$^\circ$
well.  The significance of the fit is larger than 10$\sigma$.

The measured invariant mass of the hypothetical X17 particle obtained from
the fit is: 16.94 $\pm$ 0.47~MeV(stat)/c$^2$ and the
intensity ratio of the X17 particle was found to be:

\begin{equation}
\label{eq:alpha}
\begin{split}
\frac{B_{X17}(GDR \rightarrow g.s.)}{B_{X17}(GDR \rightarrow 2_1^+)} =
  \frac{\alpha_{ground}}{1 - \alpha_{ground}} =
  0.08 \pm 0.19
\end{split}
\end{equation}

\noindent
Although the error bar is very large, it agrees within $1\sigma$ error
bar with the intensity ratio of the corresponding $\gamma$-rays of
I$_\gamma$(GDR$\rightarrow $ g.s.)/I$_\gamma$(GDR$\rightarrow
2_1^+$)=0.18$\pm$0.02 and also with the intensity ratio of
$e^+e^-$-pairs of
I$_{e^+e^-}$(GDR$\rightarrow$g.s.)/I$_{e^+e^-}$(GDR$\rightarrow$2$_1^+$)=0.25$\pm$0.03.

\section{Summary}

We reported on a new direction of X17 research. For the first time, we
successfully detect this particle in the decay of the Giant Dipole
Resonance (GDR). Since this resonance is a general property of all
nuclei, the study of GDR may extend these studies to the entire
nuclear chart.

We have studied the GDR (J$^\pi$ =1$^-$) E1-decay to the ground state
(J$^\pi$ =$0^+$) and to the first excited state (J$^\pi$=$2_1^+$) in
$^{8}$Be.  The energy-sum and the angular correlation of the $e^+e^-$
pairs produced in the $^{7}$Li($p$,e$^+$e$^-$)$^{8}$Be reaction was
measured at a proton energy of E$_p$=~4.0 MeV.  The gross features of
the angular correlation can be described well by the IPC process
following the decay of the GDR.  However, on top of the smooth,
monotonic distribution of the angular correlation of $e^+e^-$ pairs,
we observed significant anomalous excess at about 120$^\circ$ and
above 140$^\circ$.

The $e^+e^-$ excess can be well-described by the creation and
subsequent decay of the X17 particle, which we have recently suggested
\cite{kr16,kr21,kr22}. The invariant mass of the particle was measured
to be ($m_\mathrm{X}c^2 = 16.95 \pm 0.48$(stat.)~MeV), which agrees
well with our previous results.

The present observation of the X17 particle in an E1 transition
supports its vector/axial-vector character.

\

\section{Acknowledgements}

We wish to thank Z. Pintye for the mechanical and J. Molnar for the
electronic design of the experiment.  This work has been supported by
the GINOP-2.3.3-15-2016-00034 and
\noindent GINOP-2.3.3-15-2016-00005 grants.

\end{document}